# Dual-target function validation of single-particle selection from low-contrast cryo-electron micrographs

**Youdong Mao[a, *], Luis R. Castillo-Menendez[a], Joseph Sodroski[a,b,c]**

[a]Dana-Farber Cancer Institute, Department of Microbiology and Immunobiology, Harvard Medical School, Boston, MA 02115, USA. [b]Department of Immunology and Infectious Diseases, Harvard School of Public Health, Boston, MA 02115, USA. [c]Ragon Institute of MGH, MIT and Harvard, Cambridge, MA 02139, USA.

*Corresponding author. Tel: +1 617 632 4358. Fax: +1 617 632 4338. E-mail address: youdong_mao@dfci.harvard.edu (Y. Mao).

## ABSTRACT

**Weak-signal detection and single-particle selection from low-contrast micrographs of frozen hydrated biomolecules by cryo-electron microscopy (cryo-EM) presents a practical challenge. Cryo-EM image contrast degrades as the size of biomolecules of structural interest decreases. When the image contrast falls into a range where the location or presence of single particles becomes ambiguous, a need arises for objective computational approaches to detect weak signal and to select and verify particles from these low-contrast micrographs. Here we propose an objective validation scheme for low-contrast particle selection using a combination of two different target functions. In an**

implementation of this dual-target function (DTF) validation, a first target function of fast local correlation was used to select particles through template matching, followed by signal validation through a second target function of maximum likelihood. By a systematic study of simulated data, we found that such an implementation of DTF validation is capable of selecting and verifying particles from cryo-EM micrographs with a signal-to-noise ratio as low as 0.002. Importantly, we demonstrated that DTF validation can robustly evade over-fitting or reference bias from the particle-picking template, allowing true signal to emerge from amidst heavy noise in an objective fashion. The DTF approach allows efficient assembly of a large number of single-particle cryo-EM images of smaller biomolecules or specimens containing contrast-degrading agents like detergents in a semi-automatic manner.

## 1. Introduction

Image formation in electron microscopy is understood as the weak-phase approximation of thin, electron-penetrable objects (Spence, 2003). The electron image formed after the objective lens is a convolution of the exit wave function passing through the object with the point spread function of the objective lens. The phase-contrast transfer function (CTF), which is the Fourier transform of the point spread function of the objective lens, gives rise to a tradeoff between the resolution transfer and the contrast transfer (Frank 2006). The phase contrast increases with increasing defocus (underfocus); however, a higher defocus leads to greater image aberration as a result of the increase in the point spread function in single image formation. At a lower defocus, less aberration allows a better transfer of the high-frequency information of the object into the image, but gives rise to a loss of low-frequency contrast. To achieve higher resolution imaging and reconstruction of the object, data collected at both relatively low and high defocus should be employed to correct the effect of CTF (Frank, 2006; Penczek et al., 1997). Exclusion of lower defocus data can be problematic for optimal CTF correction and high-resolution reconstruction (Ludtke and Chiu, 2003).

Cryo-electron microscopy (cryo-EM) of unstained frozen hydrated biological macromolecules typically suffers from low contrast. Many factors can contribute to lower contrast in electron image formation, such as: (1) the use of low doses to preserve the structural integrity of biomolecules, (2) the use of lower defocus to include high-frequency information for optimal CTF correction, (3) the use of an objective lens of lower spherical aberration that allows a higher information limit, (4) imperfections in the detective quantum efficiency of image recording devices (CCD camera and film), (5) the specimen movement caused by either charging or mechanical perturbation, (6) the small size of the macromolecules of interest, (7) variation of the ice thickness in the cryo-specimens, and (8) the presence of detergents or heavy glycosylation on the protein surface. Therefore, improving contrast often involves tuning factors that compromise the acquisition of high-resolution information, such as using a large defocus or a higher electron dose.

Given the limitations on improving the contrast in individual cryo-EM images, it is possible to improve the contrast of averaged single-particle images by including more images in the average, which could also lead to an improvement of resolution. Therefore, improving resolution of cryo-EM structures of biomolecules often requires that more images are collected and analyzed. Selection of single-particle images from low-contrast cryo-EM micrographs represents a significant bottleneck in analyzing a large number of images. Manual selection can be very time-consuming and is prone to errors resulting from subjective variables. Computerized particle selection is therefore practically crucial for the assembly of a large number of single-particle images for cryo-EM structure refinement. The development of an objective computational procedure to select, evaluate and validate single-particle images from extremely low-contrast micrographs represents a critical prerequisite for determining higher resolution structures of smaller protein complexes.

Over the past few decades, a number of computational tools have been developed toward the goal of automatic particle identification and verification (Adiga et al., 2005; Baxter et al., 2009; Chen and Grigorieff 2007; Frank and Wagenknecht, 1984; Hall and Patwardhan, 2004; Huang and Penczek, 2004; Langlois et al., 2011; Mallick et al.,

2004; Rath and Frank 2004; Ogura and Sato, 2004; Roseman 2003, 2004; Voss et al., 2009; Wong et al., 2004; Zhao et al., 2013; Zhu et al., 2004). For example, a template-matching approach has proven to be quite efficient in automated particle picking (Rath and Frank, 2004; Roseman 2003, 2004). Recent automated particle selection approaches based on machine learning relieve the burden of post-picking manual selection (Langlois et al., 2011). It is generally thought that cross-correlation-based approaches can successfully pick particles with a signal-to-noise ratio (SNR) of 0.1 or higher from cryo-EM micrographs. It remains unclear whether these approaches can still pick particles automatically at a lower SNR and how the picked low-contrast particles can be objectively verified afterwards. In this paper, we investigate methods to select and verify particles from extremely low-contrast micrographs in an objective manner. A validation scheme using dual target functions (DTF) for identifying and detecting weak signal in single-particle micrographs is proposed and examined (Figure 1A). We quantitatively characterize the performance of DTF validation tests on simulated micrographs exhibiting a wide range of SNRs. Through comparative DTF studies, we demonstrate that the use of a second target function can robustly evade any over-fitting and reference bias incurred by the use of the first target function.

## 2. Theory

### 2.1. Problem of weak-signal detection

In the practice of cryo-EM structure determination, one first needs to identify and select single-particle images with an appropriate box size from cryo-EM micrographs that contain projection views of the same macromolecule in random orientations. Because the macromolecule may vary in thickness along different viewing directions, the image contrasts of the same structure in different viewing orientations can vary over a wide range (potentially up to ~10 times). For a given view, the local contrast of the projection image may also vary among different subunits and domains. For small protein complexes, when the overall contrast is low, certain views or certain parts of a view can be another 2-10 times lower in contrast. This contrast variation can result in substantial ambiguity in subjectively identifying projection images of small particles. Given the limitations on electron dose for imaging that preserves the fine structure, partial loss of contrast can result in the illusion that some views are smaller than expected in size or even absent. Thus, manual selection based solely on clear visibility can cause substantial subjective bias in the particle statistics, which may give rise to greater shape errors in the 3D reconstruction. As the local SNR of a projection image falls below 0.05, it becomes increasingly difficult to distinguish signal from noise by eye. In principle, weak signals that fall below the clear visibility threshold for human eyes can potentially be detected and verified by computational procedures that objectively extract signal from noise. Such approaches applied to the problem of weak-signal detection may render current cryo-EM techniques capable of reliably detecting smaller particles.

## 2.2. Over-fitting and reference bias

As noise can self-correlate to create a false-positive fit to a target function, over-fitting of noise can potentially afflict any target function or computational algorithm. This can be a barrier for weak-signal detection from high background noise. In image analysis, when an experimental noisy image is compared with a reference image, the alignment parameters of the image (displacement and rotation) can be biased by the reference. This type of over-fitting of noise is generally referred to as reference bias or model bias. However, optimization of a multi-dimensional data set against different target functions can have dramatically different effects on over-fitting or reference bias. For example, the cross-correlation function exhibits a reference dependency that can persist in many iterations of optimization (Shaikh et al., 2003; Sigworth 1998). In contrast, the maximum-likelihood (ML) approach using a log-likelihood function regularly permits an escape from reference bias (Sigworth, 1998).

In image alignment, despite the aforementioned caveats, over-fitting can be avoided by the use of a featureless template, such as a Gaussian circle, or by employing a reference-free approach. On the other hand, if the reference used in image alignment does represent the intrinsic features of the signal present in the image, over-fitting is less likely to dominate, given a sufficient SNR. For a specific target function, it is important to define the lower bound of SNR beyond which the specific target function begins to fail in detecting or aligning signal.

## 2.3. Concept of dual target function (DTF) validation

In dealing with the problem of weak-signal detection, over-fitting and reference bias in a single target function can certainly blur the "boundary" between signal and noise, creating a barrier for true signal to stand out. Nevertheless, it is mathematically prohibited that, under the same set of fitting parameters, the over-fitting of noise to one specific target function will be optimally reproduced by another target function that is not equivalent to, or correlated with, the first function. Thus, the conceptual foundation of DTF validation lies in an appropriate choice and use of a second target function that significantly differs from the first one; employing such a second target function should remove any potential over-fitting of noise resulting from the use of the first target function, allowing the true signal to be recovered. This DTF strategy can be used to detect and verify the weak signal present in cryo-EM micrographs.

Computerized procedures for weak-signal detection in single-particle cryo-EM involve two steps: particle picking and particle verification. A number of algorithms have been developed to automate template-matching procedures for particle picking; these procedures require subsequent manual selection of particles, in some cases with the help of data clustering to expedite the rejection of false positives (Hrabe et al, 2012; Shaikh et al., 2008; Zhao et al., 2013). The majority of algorithms implementing template matching for particle-picking applications are based on the cross-correlation function, which calculates the normalized correlation between the template image and a local area of a micrograph. A disadvantage of the cross-correlation function is its sensitivity to noise, which can create false correlation peaks that do not result from real signal. However, these false, noise-based peaks of cross-correlation still retain the intrinsic statistical properties of noise; that is, their appearance in the 2D positions of a correlation map is random. When these pure noise images that are boxed out of a micrograph are aligned against a different target function, such as a maximum-likelihood (ML) estimator, the similarity of images indicated by the false correlation peak cannot be reproduced, due to the random nature of noise.

In the presence of signal and the absence of noise, the cross-correlation function and ML estimator both lead to the same solution for the image alignment problem (Sigworth 1998; Sigworth et al., 2010). However, in the presence of noise, the cross-correlation function demonstrates an increasing propensity to identify false-positive particles as the SNR decreases (Glaeser 2004; Zhu et al., 2004). In principle, although the ML estimator does not absolutely exclude the occurrence of false positives, its exhaustive probability search across parameter space substantially reduces the effect of false positives over the iterations of the expectation-maximization algorithm (Sigworth 1998). Therefore, following initial particle picking, particle verification by a reference-free ML alignment can be implemented (Figure 1); the generation of a clear 2D structure in the class averages, particularly if this structure is consistent with other available data, is strong evidence of the alignment of real signal in the images. When using reference-free alignment or using a featureless Gaussian circle as an initial reference, the imaging noise or false positives cannot dominate the ML optimization in the presence of sufficient signal. Therefore, an important question to be answered quantitatively in this study is, "What level of SNR is sufficient to permit DTF validation to succeed?".

## 3. Methods

### 3.1. A practical implementation of the DTF validation procedure

Throughout this study, the following implementation of DTF validation was applied to 26 data sets of either pure noise or simulated low-contrast micrographs of the trimeric ectodomain of the influenza hemagglutinin (HA) glycoprotein (Weis et al., 1990). An illustration of the DTF validation procedure is summarized in Fig. 1B.

Step 1: Particle picking by fast local cross-correlation. We used template matching by fast local cross-correlation implemented in SPIDER to pick particles (Frank et al, 1996). The SPIDER script, lfc_pick.spi, has been studied in the case of the ribosome (Rath and Frank, 2004) and has served as a control for the recent development of a reference-free particle-picking approach (Langlois et al, 2011). This procedure applies a fast local correlation (FLC) function to particle recognition, following Roseman's (2003) approach. In our study, we picked particles using single 2D templates, as described in the specific experiments below. Note that previous studies have shown that using the FLC function with a single template can pick many views of particles (Rath and Frank, 2004).

Nonetheless, it has been suggested that using more templates can potentially reduce the number of false positives that are picked (Roseman 2003, 2004; Rath and Frank, 2004; Glaeser 2004).

Step 2: Candidate particle selection by the use of a threshold in the ranking of correlation peaks and manual rejection of obvious artifacts. The SPIDER particle-picking program (lfc_pick.spi) sorts and ranks the picked particles according to their correlation peaks, from high to low peak values. Upon sorting and ranking, the potential true particles often appear at higher correlation peak values and the pure noise images at lower correlation peaks. A threshold that approximately demarcates the boundary between the potential true particles and pure noise can be used to select the initial candidate particles, followed by manual inspection of each particle and rejection of obvious artifacts. The rejection of suspected artifacts and false positives can be done in a batch mode if the picked particles are clustered into groups (for example, by multivariate statistical analysis) (Hrabe et al, 2012; Shaikh et al., 2008; Zhao et al., 2013).

Step 3: Particle validation by a reference-free ML alignment with single or multiple classes (Scheres et al., 2005; Scheres 2010). The ML-based approach for image alignment has been previously demonstrated to be quite resistant to reference bias after a sufficient number of iterations of optimization (Sigworth 1998). Image similarity measured by probability and subsequent class averages calculated by integration over all different probabilities are more sensitive to the presence of true signal (Scheres et al., 2005).  One would expect that any bias in particle selection would not persist through a number of iterations of ML alignment in a reference-free manner or using a Gaussian circle as a starting reference. In the studies below, we specifically test the ability of ML alignment to extract signal from noisy images and to remove reference bias that was introduced by template matching..

### 3.2. Simulation and DTF testing of noise micrographs

We first simulated 200 micrographs of only Gaussian noise by the SPIDER command MO (option R with Gaussian distribution). Each micrograph has dimensions of 4096 x 4096 pixels. We then used one projection view of the ~11-Å human immunodeficiency virus (HIV-1) envelope glycoprotein trimer (Mao et al., 2012) as a template for particle picking from the simulated Gaussian-noise micrographs. The box size is 256 x 256 pixels. In each micrograph, about 20-25 boxed images of the highest local correlation peaks were selected to assemble a particle stack of 4485 images. After particle picking and selection, each particle image was scaled 4 times to 64 x 64 pixels (using xmipp_scale) and normalized (using xmipp_normalize) (Sorzano et al., 2004). Subsequent ML alignment of a single class (using xmipp_ml_align2d) was repeated with three different starting references: (1) a noise image randomly chosen from the whole image stack; (2) a Gaussian circle; (3) an average of a random subset of the unaligned images that replicates the template used for particle picking.

To repeat the above DTF test on real experimental ice noise, we imaged a cryo-grid that was composed only of buffer solution and contained no protein sample. The composition of the buffer was 20 mM Tris-HCl, pH 7.4, 300 mM NaCl and 0.01% Cymal-6. This was the same buffer used for maintaining the HIV-1 membrane envelope glycoprotein trimer in solution during the cryo-EM data collection for its structural analysis (Mao et al., 2012; Mao et al., 2013). The cryo-grid was made from a C-flat holey carbon grid by FEI Vitrobot Mark IV. The data were collected on an FEI Tecnai G2 F20 microscope operating at 120 kV, with a Gatan Ultrascan 4096 x 4096 pixel CCD camera, at a nominal magnification of 80,000. From about 600 micrographs collected in one cryo-EM session, 218 micrographs of pure ice noise were chosen. The same particle-picking procedure performed with the simulated Gaussian noise micrographs (see above) was applied to the experimental ice noise micrographs, with the same HIV-1 envelope glycoprotein trimer template. After particle picking, the apparent ice-crystal contaminants were manually rejected from the particle set, leaving only images from amorphous ice noise. By selecting only about 10-25 boxed images of the highest local correlation peaks from each micrograph, a particle stack of 4591 images was assembled and was subjected to the same ML alignment as described above for the data from the simulated Gaussian noise micrographs. These DTF tests on both the simulated and experimental pure noise micrographs (Fig. 2) serve as controls for the subsequent examination of the effect of SNR on the success rate of DTF validation.

### 3.3. Simulation and DTF testing of low-contrast micrographs

We simulated 120 micrographs of noiseless particles corresponding to the crystal structure of the influenza A virus hemagglutinin (HA) glycoprotein ectodomain (PDB ID: 3HMG) (using xmipp_phantom_create_micrograph) (Weis et al., 1990). The simulation assumes an acceleration voltage of 120 kV, a defocus of -1 μm, a pixel size of 1.0 angstrom, and micrograph dimensions of 4096 x 4096 pixels. In each simulated micrograph, there are 323 HA molecules that assume random orientations. To add different levels of Gaussian noise to the noiseless micrographs, the standard deviation of the background of each micrograph was calculated and used as input to simulate a background Gaussian noise image that was added to the noiseless micrographs. This results in micrographs with Gaussian noise added to yield SNRs of 0.1, 0.05, 0.02, 0.01, 0.005, 0.002, 0.001 or 0.0005. A typical series of a simulated noiseless micrograph and the derived noisy micrographs at different SNRs is shown in Fig. 3.

For the simulated micrographs at each SNR value, we conducted DTF tests using three different templates for particle picking, i.e., a Gaussian circle, one projection view of the influenza virus HA trimer filtered to 30 Angstroms, and one projection view of the HIV-1 envelope glycoprotein trimer filtered to 30 Angstroms (Fig. 5). Each set of micrographs with a given SNR, which is selected by a particular particle-picking template, is treated as a separate case. Therefore, there are 8 x 3 = 24 cases studied and compared in our DTF tests. For each case, a stack of 38,760 particle images was assembled, based on a selection threshold of 323, from 120 simulated micrographs. The original box dimension for particle picking was 180 x 180 pixels. After particle picking and selection, each particle image was first scaled 3 times to a dimension of 60 x 60 pixels and normalized for the background noise, then subjected to unsupervised multi-reference ML classification into 5 classes.

## 4. Results

### 4.1. DTF tests on simulated and experimental noise

As a control experiment to investigate the ability of the DTF approach to resist reference bias, we conducted DTF tests on simulated micrographs that contain only Gaussian noise. A single 2D projection of the HIV-1 envelope glycoprotein trimer was used as a template for picking "particles" by FLC (Target Function A) (Fig. 2A). Images with the highest local correlation peaks were selected and subjected to ML alignment, using three different starting references for ML optimization (Target Function B). In the first DTF test, a raw pure noise image randomly chosen from the particle stack was used as a starting reference for ML optimization (Fig. 2B). Over more than 3000 iterations of ML alignment, no 2D structure resembling the particle-picking template was observed. The resulting average image in each iteration was still a random noise image. We then used a Gaussian circle as the starting reference to repeat the ML optimization (Fig. 2C). Again, the resulting average image contained only random noise but no observable 2D model. As a third starting reference for ML optimization, we used the average of template-selected particle images without any further alignment. Notably, this average closely resembles the HIV-1 envelope glycoprotein template used for particle picking (Fig. 2D), and apparently results from reference bias in template-based particle picking by the FLC target function. Using this average image as a starting reference for the ML alignment, the replica of the template fades out in the average image and disappears upon the convergence of ML optimization. Thus, the DTF approach can remove reference bias associated with the alignment of pure noise during the particle-picking process.

Next, we asked if the results observed with the simulated micrographs of Gaussian noise would be reproduced with images of actual cryo-EM noise resulting from amorphous ice. We repeated the aforementioned DTF tests on the data set assembled from experimental ice noise micrographs.  As shown in Fig. 2E-G, when aligned by ML, no structure was observed after more than 3000 iterations of optimization no matter what type of starting reference was used. In all three cases, the converged class average in ML showed a blank image without any observable signal. Therefore, images of experimental ice noise taken by a CCD camera reproduce the results seen for simulated Gaussian noise, supporting the notion that the experimental cryo-EM noise from amorphous ice basically exhibits Gaussian-like behavior (Frank, 2006).

**4.2. The simulated low-contrast micrographs**

Next, we tested the FLC-based particle-picking program on a number of simulated micrograph sets. Different levels of Gaussian noise were added to the same simulated noiseless micrographs, each containing 323 particles of influenza virus HA trimers in random orientations, to create images with SNRs of 0.1, 0.05, 0.02, 0.01, 0.005, 0.002, 0.001 and 0.0005. Figure 3 shows a typical noiseless micrograph (Fig. 3A) and the micrographs with different SNRs derived from it (Fig. 3B-H). As expected, the visibility of particles is drastically diminished in the lower SNR ranges. Because the loss of visibility creates difficulty in directly verifying the false and true positives in the same low-contrast micrograph in our particle-picking test, the original noiseless micrograph from which the low-contrast micrograph was derived can be used to verify particle-picking performance (Fig. 4).

Using the noisy micrographs containing the randomly oriented influenza virus HA trimers, we repeated the particle-picking tests with three different templates (a Gaussian circle, one projection view of the influenza virus HA trimer, and one projection view of the HIV-1 envelope glycoprotein trimer). Figures 5A-C show the plots of the correlation peak versus the rank number of picked particles. Notably, when the Gaussian circle was used as a template (Fig. 5A), the plots corresponding to SNRs of 0.1, 0.05, 0.02, 0.01 and 0.005 all show a clearcut drop-off in the value of the correlation peak at a rank of 323, the number of particles simulated in each micrograph (Frank and Wagenknecht, 1984). All of these 323 picked particles with high correlation peak values were confirmed to be true positives (Fig. 6). When the Gaussian circle was used to pick particles from the micrograph with an SNR of 0.002, the plot of the correlation peaks still exhibited a discernible drop-off at N = 323, but with a much smoother edge (Fig. 5A). The drop-offs in correlation peak values are smoother and less prominent at lower SNR values (0.001 and 0.0005). Using 323 as the threshold for particle selection, the number of false positives increased to approximately 2% at an SNR of 0.001, and to approximately 7% at an SNR of 0.0005 (Fig. 5D). These false-positive rates are surprisingly low, given the very low values of the corresponding SNRs.

We evaluated the specificity of particle picking when using templates other than a Gaussian circle; i.e., one projection view of the influenza virus HA trimer itself and one projection view of the HIV-1 envelope glycoprotein trimer, which bears little similarity to the HA trimer (Fig. 5B and C). For both templates, clear drop-offs in the correlation peak-ranking plots at N = 323 were observed at SNR values of 0.005 and higher. Notably, in all cases of using different templates in the particle-picking test, there were no false positives at SNR values greater than or equal to 0.005 (Figs. 5D and 6A-C). However, using the Gaussian circle template allowed better centering of picked particles than using the other two templates. Among the cases compared here, the centering of picked particles was the worst when the dissimilar 2D structure (the HIV-1 envelope glycoprotein trimer) was used as a template for micrographs with the lowest SNR (0.0005) (Fig. 6F). Apparently, particle recognition is less sensitive to the detailed shape of the particle-picking template than are the specificity and particle-centering accuracy.

Thus, the use of a dissimilar template succeeded in particle recognition at large, but resulted in a greater mis-centering of the picked particles and more false positives at the lowest SNRs (0.002, 0.001 and 0.0005) (Figs. 5D and 6).

**4.3. DTF tests on the low-SNR particle sets**

We evaluated the ability of the DTF approach to verify the presence of signal in the particles selected from micrographs with different SNRs by different particle-picking templates. Using a threshold of 323 to select the particles with higher correlation peaks, we subjected the selected particles to multi-reference ML classification and averaging (Fig. 7). The particle sets selected from micrographs with different SNRs using different templates were treated and classified separately, and the results were compared among the different SNRs and different particle-picking templates. Strikingly, after ML optimization, the class averages all recapitulated the projection views of the influenza virus HA trimer, no matter what type of particle-picking template was used, for those data sets derived from micrographs with SNRs higher than 0.001. The ML optimization results using particles selected from micrographs with SNR = 0.002 were comparable for those selected by the Gaussian circle template (Fig. 7D) and those selected by the dissimilar HIV-1 envelope glycoprotein trimer template (Fig. 7F). Evidently, the model used for the particle-picking template does not govern the outcomes of ML optimization when sufficient signal is present.

Of note, the DTF test intermittently succeeded in aligning true signal even at an SNR as low as 0.0005. Nonetheless, at low SNR values, the frequency of such successful alignments and the quality of the class averages produced dropped significantly, as expected. Thus, at the lowest SNRs (0.001 and 0.0005), DTF validation became inefficient in verifying signal for this data set of 38760 particles. Considering that an SNR of 0.001 is unusually low and often can be avoided experimentally, the DTF tests on the simulated low-contrast micrographs should be relevant to the analysis of real cryo-EM experimental data.

**4.4. Effect of reference bias in particle selection and its limitations**

The fitting parameters in the particle-picking problem are the X-Y coordinates of the particle box. The choice of template in particle picking appears to bias the coordinates of the boxes. As shown in Fig. 6, the selected particles were best centered when using the Gaussian circle as a template, whereas the particle boxes deviated most from the particle centers when the template was one projection view of the HIV-1 envelope trimer, a template that does not reflect the intrinsic structures in the micrographs. Consequently, the average image of the picked particles after boxing and before alignment closely resembled the template image (See the columns with the starting references (S. Ref.) in Fig. 7). However, the template neither changes the true signal in the boxed particle images nor is used in signal alignment by the ML estimator, allowing objective signal validation by the second target function. As seen in all the ML optimization tests on the particle sets selected from micrographs with different SNRs, upon convergence, the class averages either show the projection views of the influenza virus HA trimer (if successful) or show a blank noise image (if failed) (See the columns showing the 1000$^{th}$ iteration in Fig. 7). At an SNR of 0.002 and higher (Fig. 7A-F), the converged ML class averages all clearly recover the projection views of the influenza virus HA trimer. At SNRs of 0.001 and 0.0005 (Fig. 7G-L), there is still partial success in recovering the projection views of the influenza virus HA trimer by ML optimization; however, more than half of the class averages at the lowest SNR tested (0.0005) (Fig. 7J-L) become a blank noise image, indicating a failure in signal detection. These results are consistent with those obtained with images of pure noise (Fig. 2). In no case does the converged ML class average recapitulate the particle-picking template. Thus, reference bias from the FLC function during particle selection can be removed by the ML function, allowing true signal to emerge in spite of extremely low SNRs.

## 5. Discussion

### 5.1. Low-SNR performance

Cryo-EM structure determination of progressively smaller biomolecular complexes necessitates picking and verifying particles from low-SNR micrographs. The risks of reference bias and the introduction of noise into the structure increase at low SNR levels. The DTF approach attempts to guard against these pitfalls. The control experiments with simulated micrographs of Gaussian noise demonstrated that the reference bias derived from the FLC function does not translate into reference bias in the ML function, in either reference-free or reference-based alignment. This conclusion also applies to the alignment of experimental cryo-EM ice noise. Together, these control experiments lay the rational foundation for the DTF validation of weak signals in low-contrast micrographs.

The DTF validation tests presented in this study make a number of critical points. First, the reference bias resulting from the FLC-based particle picking can be fully removed by the ML-based alignment performed in a reference-free manner or using a Gaussian circle as the starting reference. Second, it is impractical to pick particles manually from micrographs with SNRs between 0.0005 and 0.01. However, the FLC implementation in SPIDER successfully picks particles with SNRs as low as 0.0005. Together with previous studies (Roseman 2003, 2004; Rath and Frank, 2004), our results suggest that the FLC approach is highly sensitive to the presence of very weak signal. Third, the typical DTF implementation suggested in this study, that is, the combination of FLC and ML evaluation, provides a highly sensitive, objective way to detect and validate signal from extremely low-SNR micrographs, even though the particles in the micrographs may not be visually obvious.

**5.2. Differences between FLC and the projection-matching algorithm**

The requirements for template matching in the particle-picking process differ somewhat from those for projection matching in structure refinement. In projection matching, one needs to be able to detect the specific features that distinguish one projection view from another. The calculation of a cross-correlation in projection matching generally involves two images of similar dimensions. In the particle-picking problem, one aims to detect the general presence of particles regardless of the detailed structure of each particle. In FLC calculations, the local correlation may be among two images of different dimensions. Therefore, fast template matching in particle picking needs only to calculate a low-frequency correlation in Fourier space in a coarse-grained manner (Roseman 2003). This property renders the performance of FLC-based particle picking relatively insensitive to changes in the specific shape of the template. Quantitative differences between the two approaches have been discussed previously (Roseman 2003). In our study, we found that the use of a dissimilar structure as the particle-picking template only marginally increased the number of false positives. As a result, a Gaussian circle may be a preferred picking template in the initial stage of automated particle picking, thus avoiding any potential selection bias (Glaeser, 2004). Once a data set has been vetted by DTF and other validation approaches, it should be feasible to use the initial reconstruction from the data set to repeat the particle picking with multiple templates that more closely resemble the structure in the data set (Glaeser 2004; Hrabe et al., 2012; Zhao et al., 2013). This re-iteration of particle picking and re-assembly of the particle data set potentially can recover a majority of the false negatives from the early phase of particle selection.

**5.3. Caveats in the application of DTF to experimental data**

Our quantitative characterization of the capabilities of the DTF test studied ideal cases with synthetic data. Differences exist between simulated and real micrographs in both the particle and the noise components. Our simulated particles are homogeneous, whereas real particles may exhibit heterogeneity in conformation, beam-induced movement, defocus values, local ice thickness and sample charging, among others. Our simulated low-contrast micrographs are free of ice contaminants, which are found to some extent in experimental cryo-EM micrographs. As the false positives derived from ice contaminants often have high correlation peaks, they can appear in the micrographs at a wide range of SNRs. Additionally, the background ice noise may also deviate from a strict Gaussian distribution. Thus, the application of the DTF approach to actual experimental cryo-EM micrographs may deviate from the simulated ideal behavior (Frank 1984; Rath and Frank, 2004). For example, the degree of the drop-off in the correlation peak-ranking plot may be less than ideal, or the level of DTF efficiency at different SNRs may be reduced by the above-mentioned heterogeneity in particles and/or background. Despite these hypothetical differences between real and ideal experiments, the mathematical principle behind DTF validation remains true, i.e., any over-fitting by the first target function (FLC) in particle picking can be removed by the second target function (ML) in signal alignment.

Several additional issues should be considered when applying the DTF approach to experimental data. First, ice contaminants are the most frequent false positives in FLC particle picking. Recent advances in applying machine learning to particle selection can largely remove these types of false positives, with little manual intervention (Langlois et al., 2011). Moreover, it is often straightforward to remove ice contaminants manually. Second, the selection threshold (N) representing the number of true-positive particles is not precisely known in real experiments. However, the experimental N can be approximately estimated from the protein densities in the hole of the supporting carbon film. Third, experimental SNR is expected to fluctuate, in contrast to the fixed SNR used in our simulation studies. Therefore, image background normalization could increase the sensitivity in detecting weak signals.

Note that the SNR calculated for a whole micrograph is often lower than the SNR calculated from boxed single-particle images, given that there are more empty background areas in the micrograph than in appropriately boxed single-particle images. When extrapolating the results of this study to the SNR of single-particle images, the SNR of a whole micrograph should be multiplied by a factor of 2 to be equivalent to the SNR of boxed particle images.

## 5.4. False positives

Although false-positive particles will inevitably be picked by the cross-correlation function, the percentage of false positives in the candidate particle pools can be reduced by manual curation on both an individual particle level and a class-average level (Rath and Frank 2004; Roseman 2004; Shaikh et al., 2003; Hrabe et al., 2012). A reference-free ML alignment that leads to a clear 2D structure in class averages should allow an unambiguous distinction between weak signal and strong noise. Under conditions of reference-free ML alignment, the false positives from pure noise cannot dominate the image alignment. Instead, through unsupervised alignment by ML, it should be possible to restore the weak signal in the presence of a small fraction of false positives in the data set.

Removing all false positives will be unlikely in real experiments involving a very large data assembly in that the appropriate selection threshold is not known and may vary from micrograph to micrograph. If a drop-off is observed in the correlation-peak ranking plot, the threshold can be estimated from the ranking number where the drop-off occurs (Frank and Wagenknecht, 1984). However, in real cryo-EM micrographs, there are often more or less ice contaminants or non-particle features, which may be picked and become false positives. These non-particle features often have stronger correlation peaks and are readily recognizable and can be manually rejected from the data set (Rath and Frank, 2004).

**6. Conclusion**

In this work, we examined the ability of the dual-target function (DTF) approach to select and validate particles from highly noisy micrographs over an SNR range where manual particle picking becomes impractical. We characterized the quantitative performance of FLC-based particle selection and ML-based particle verification over a wide range of SNRs. The DTF validation approach, which combines the two target functions, represents a sensitive, objective way to assemble particles for downstream cryo-EM structure refinement. Importantly, the DTF approach does not transfer any reference bias from the FLC target function to the ML target function. This makes possible the robust detection and objective validation of weak signal. We also quantitatively characterized the critical SNR where DTF performance begins to degrade. We found that the critical SNR is surprisingly small, as low as 0.001, given the size of the data set (38760 particles) tested in each case. This study suggests that it is possible to select particles automatically or semi-automatically from extremely noisy micrographs taken at a lower defocus, or from cryo-specimens composed of smaller complexes or membrane protein complexes surrounded by contrast-degrading detergents. Looking forward, there could be alternative implementations of DTF validation, as long as the two chosen target functions are not mathematically equivalent or correlated. For example, a regularized likelihood function may provide improved sensitivity in verifying heterogeneous particles (Scheres, 2012). Improved implementation of DTF validation might further push the envelope of detecting weak signal that is difficult to ascertain subjectively.

**Acknowledgements**

The authors thank J. Jackson and B. Richter for assistance in maintaining the high-performance computing system; Y. McLaughlin and E. Carpelan for assistance in manuscript preparation. The experiments and data processing were performed in part at the Center for Nanoscale Systems at Harvard University, a member of the National Nanotechnology Infrastructure Network (NNIN), which is supported by the National Science Foundation under NSF award no. ECS-0335765. This work was funded by the National Institutes of Health (NIH) (AI93256, AI67854, AI100645 and AI24755), by an Innovation Award and a Fellowship Award from the Ragon Institute of MGH, MIT and Harvard, by the International AIDS Vaccine Initiative, and by gifts from Mr. and Mrs. Daniel J. Sullivan, Jr. and the late William F. McCarty-Cooper.

**Figure Legends**

**Figure 1**. Strategy and implementation of DTF validation. (A) The concept of DTF validation involves the use of two different target functions. The first target function deals with particle detection and the second target function with particle verification. (B) One implementation of DTF validation that is proposed in this study combines fast local correlation (FLC) and maximum likelihood (ML) target functions, which are not mathematically equivalent or correlated. User-determined templates/references are shown in the dashed boxes, designated by the terms that will be used throughout this manuscript.

**Figure 2**. The DTF results for pure noise data, both simulated and experimental. (A) A schematic flow diagram shows that "particles" were picked by FLC from pure-noise micrographs, using a single projection of the HIV-1 envelope glycoprotein trimer as a template. The picked particles were subjected to ML alignment, using different starting references. (B-D) The FLC-picked particle set, derived from the simulated Gaussian-noise micrographs, was aligned by ML, starting from a noise image randomly chosen from the particle set (B), a Gaussian circle (C), or the average of the picked particles without any further alignment (D). This starting reference for ML optimization is shown in the first column. Each row shows the history of the ML-aligned class averages at the indicated iterations of optimization ($1^{st}$ – $3000^{th}$ iteration), ending with the converged class average in the far right column. (E-G) The FLC-picked particle set, derived from the experimental ice noise micrographs, was aligned by ML, starting from a noise image randomly chosen from the particle set (E), a Gaussian circle (F), or the average of the picked particles without any further alignment (G). The averages shown in (D) and (G) appear as an FLC-generated replicate of the 2D template used in the particle picking.

**Figure 3**. The simulated micrographs with different SNRs. (A) An example is shown of a simulated noiseless micrograph containing projection views of the influenza virus HA trimers in random orientations. (B-H) A different level of Gaussian noise was added to the noiseless micrograph shown in (A) to simulate noisy micrographs at an SNR of 0.05 (B), 0.02 (C), 0.01 (D), 0.005 (E), 0.002 (F), 0.001 (G), and 0.0005 (H).

**Figure 4**. An example of FLC-based particle picking from extremely low-contrast micrographs of the influenza virus HA trimer. (A) The simulated noisy micrograph of influenza virus HA trimers at an SNR of 0.005 is shown, superposed with all 323 particle boxes (red) picked by FLC with the Gaussian circle particle-picking template. (B) The simulated noiseless micrograph that was used to derive the micrograph shown in (A), with the same 323 particle boxes (red) superposed on the micrograph. This was used for visual verification of the performance of the FLC-based particle picking, showing the absence of false positives. (C) The simulated noisy micrograph of influenza virus HA trimers at an SNR of 0.0005, superposed with all 323 particle boxes (red) picked by FLC with the Gaussian circle particle-picking template. (D) Verification of the particle-picking results in (C) on the simulated noiseless micrograph. The low contrast of particles in (A) and (C) would render manual particle picking challenging and impractical.

**Figure 5**. The correlation peak-ranking plots and differentiation of true-positive and false-positive particles in FLC-based automated particle picking. (A-C) The correlation peak-ranking plots corresponding to different SNRs, using three different particle-picking templates: (A) a Gaussian circle, (B) one projection view of the influenza virus HA trimer, and (C) one projection view of the HIV-1 envelope glycoprotein trimer. The particle-picking templates are shown in the insets. All plots are from the noisy particle micrographs derived from the same simulated noiseless micrograph of the influenza virus HA trimer. Note that the position of the drop-off in correlation peak values corresponds to 323, the number of actual influenza virus HA trimers in the simulated micrographs. (D) The plots of false positives in particle picking by the three different templates are shown, indicating that the specificity of FLC particle picking is highly dependent on the SNR and is also affected by the choice of the 2D template.

**Figure 6**. Comparison of the FLC-based particle picking results at different levels of SNR and with different templates. In each left panel, a gallery of 323 noisy particles boxed out of the influenza virus HA-containing micrographs with SNRs of 0.005 (A-C) and 0.0005 (D-F) are shown. Each right panel shows a gallery of noiseless particles picked out of the original noiseless micrograph, using the same boxing parameters and in the same sequence as in the corresponding left panel. This comparison provides a visual verification of the particle-picking performance. The particle-picking templates were a Gaussian circle (A and D), one projection view of the influenza virus HA trimer (B and E) and one projection view of the HIV-1 envelope glycoprotein trimer (C and F).

**Figure 7**. Effects of the particle-picking template used in FLC and the micrograph SNR on ML optimization. Noisy micrographs containing the influenza virus HA trimers with different SNRs were subjected to DTF testing, using different templates for particle picking. The corresponding SNRs of the micrographs from which the particle sets were picked are 0.005 (A, B and C), 0.002 (D, E and F), 0.001 (G, H and I) and 0.0005 (J, K and L). The templates used in particle picking were a Gaussian circle (A, D, G and J), one projection view of the influenza virus HA trimer (B, E, H and K) and one projection view of the HIV-1 envelope glycoprotein trimer (C, F, I and L). The particles picked by FLC were randomly divided into five classes and averaged; these "class averages" are shown in the leftmost column of each panel A-L. Using the random class averages as starting references, each assembly of data sets was subjected to multi-reference ML classification. In each panel, five rows of image series correspond to five classes generated by ML, with the class averages of the milestone iterations (1st, 10th, 50th, 100th and 1000th) shown in a row. The DTF testing results show that ML optimization can recover the weak signal of the influenza virus HA trimer if there is sufficient SNR in the images. At low SNR, ML optimization either recovered the true signal or failed, but never reproduced the template used for particle picking by FLC.

**A**

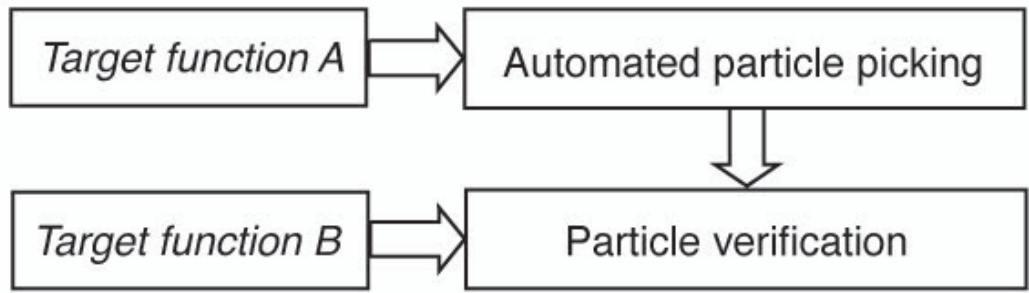

**B**

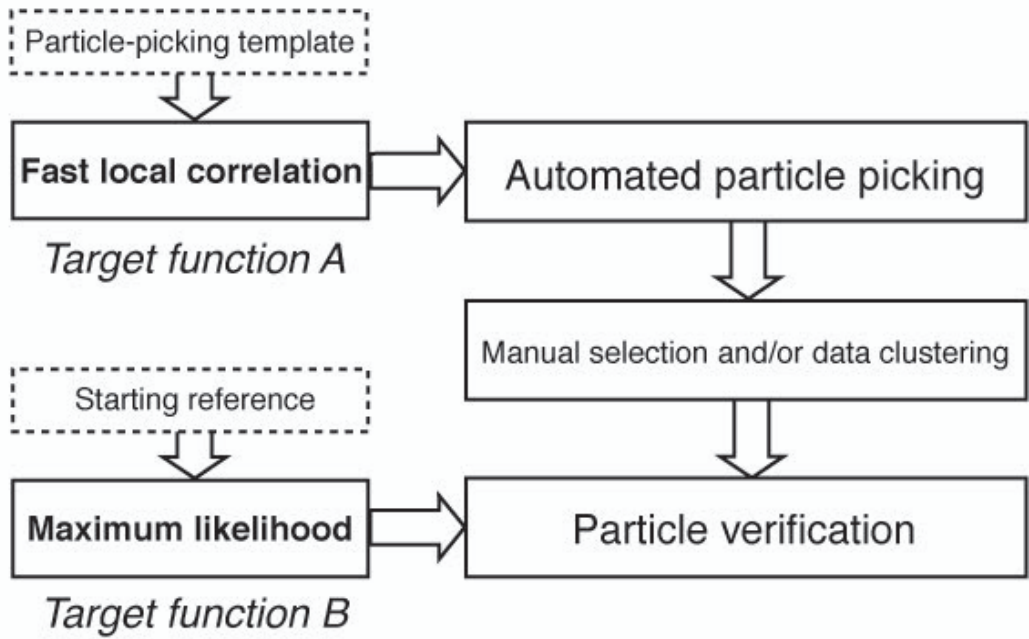

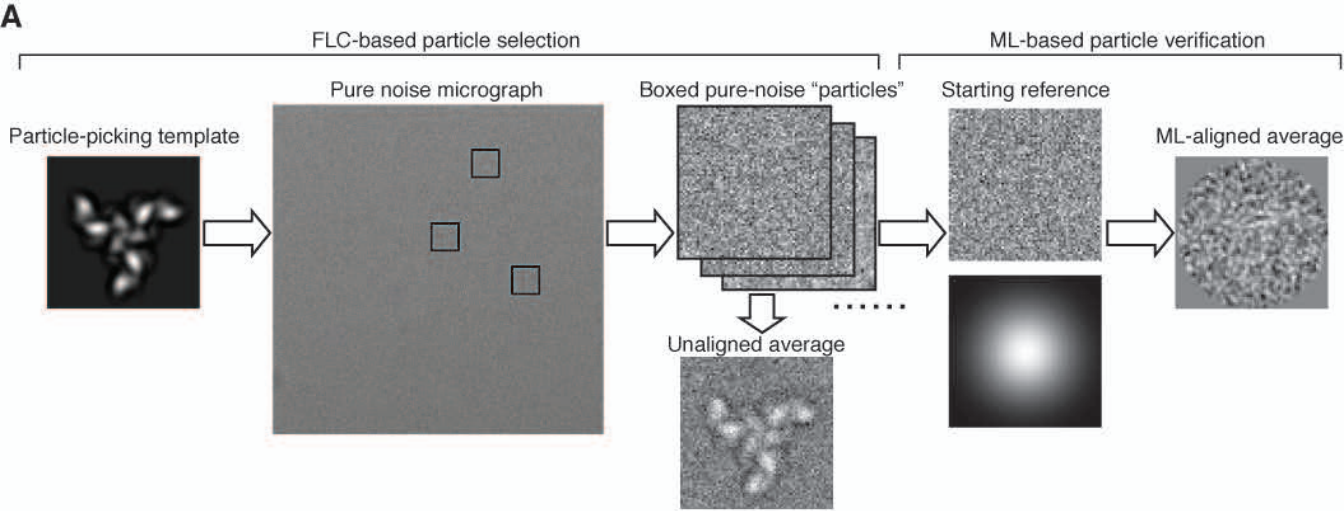

**A**

FLC-based particle selection | ML-based particle verification

Particle-picking template

Pure noise micrograph

Boxed pure-noise "particles"

Starting reference

ML-aligned average

Unaligned average

**B**

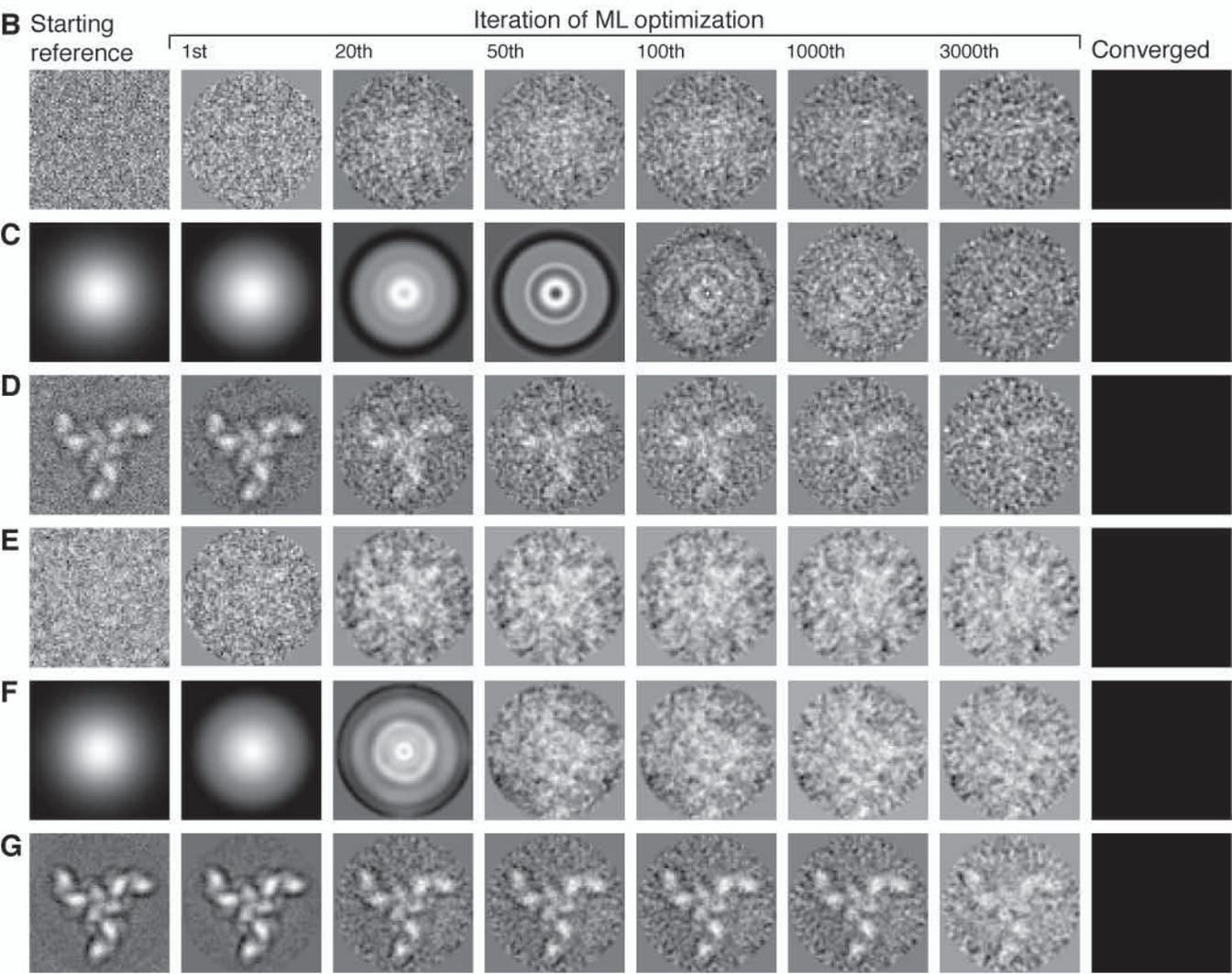

Starting reference | Iteration of ML optimization | Converged

1st | 20th | 50th | 100th | 1000th | 3000th

**C**

**D**

**E**

**F**

**G**

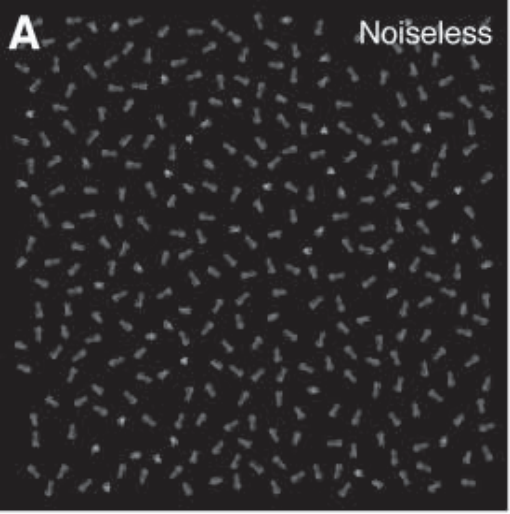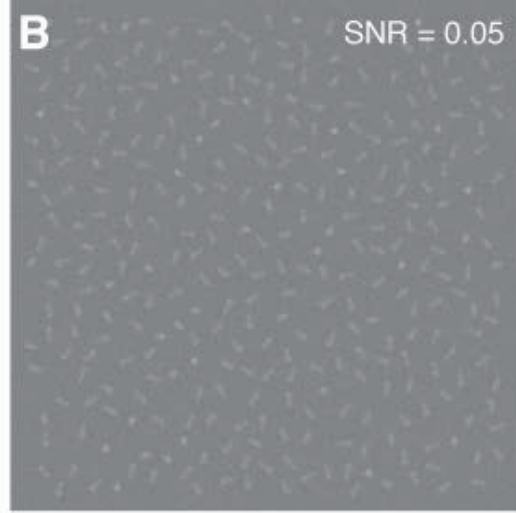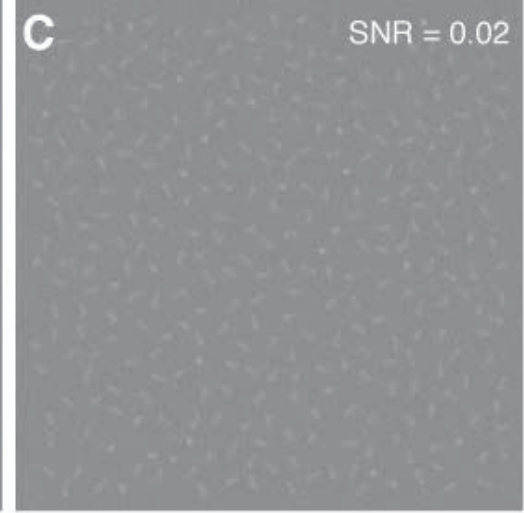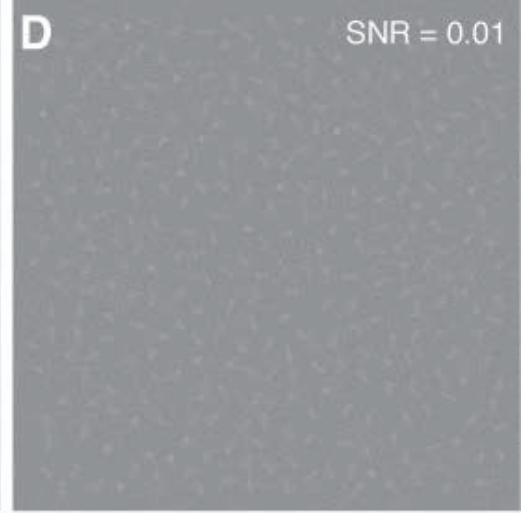
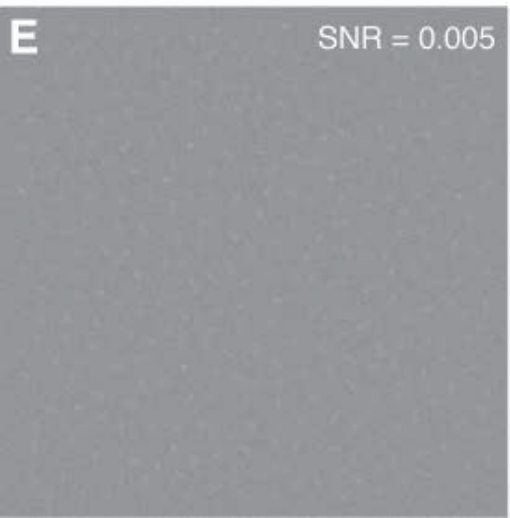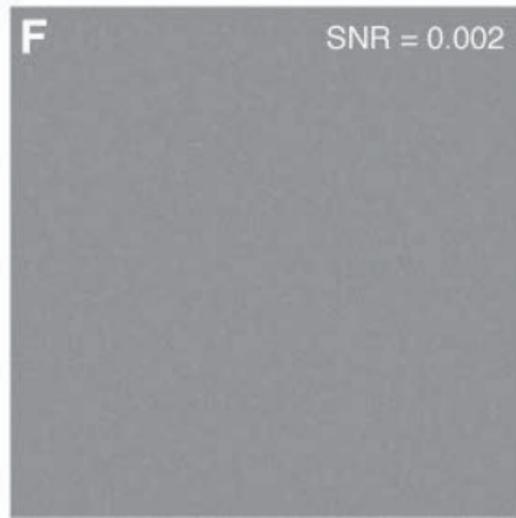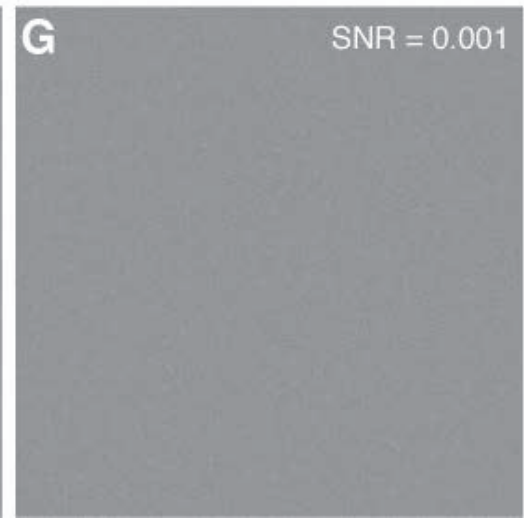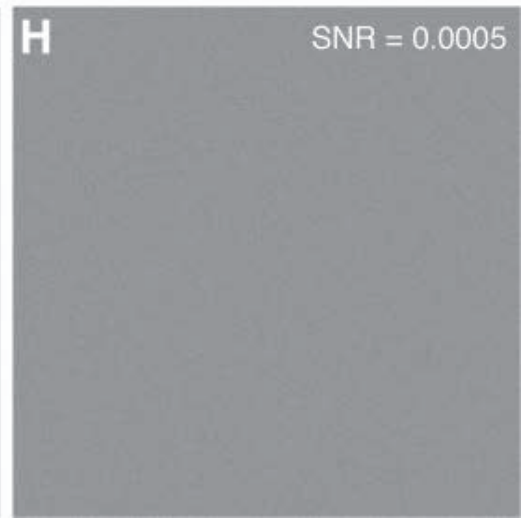

**A** Blind Test, SNR = 0.005

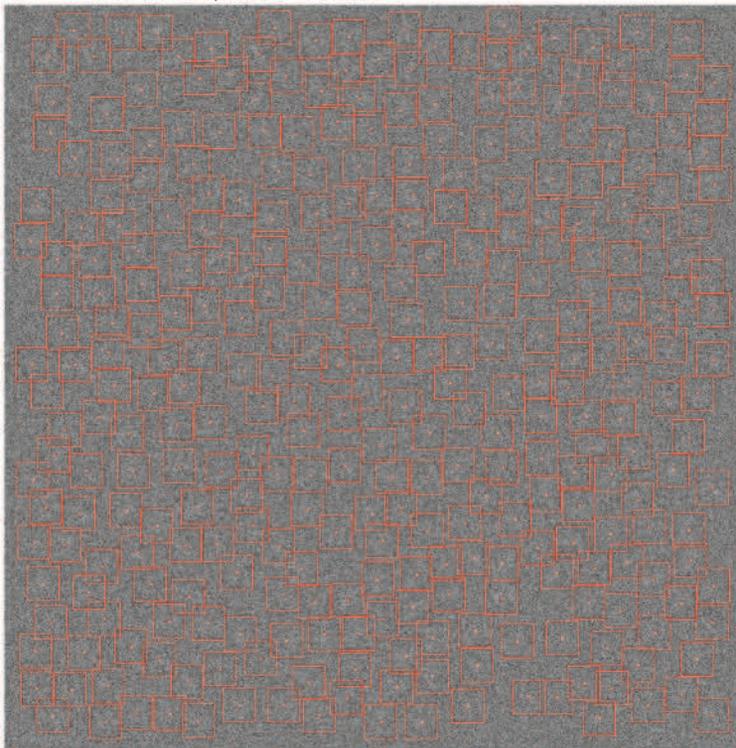

**B** Verification

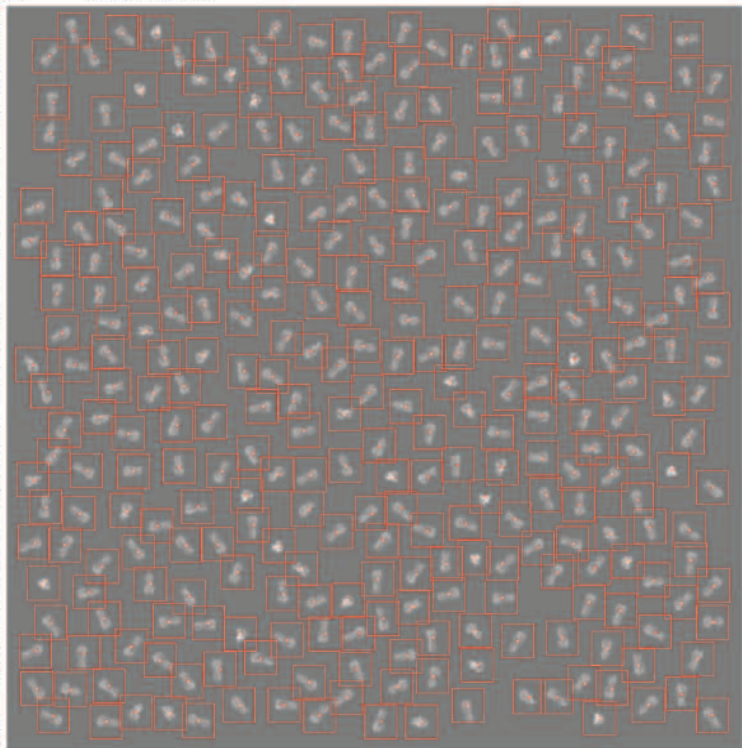

**C** Blind Test, SNR = 0.0005

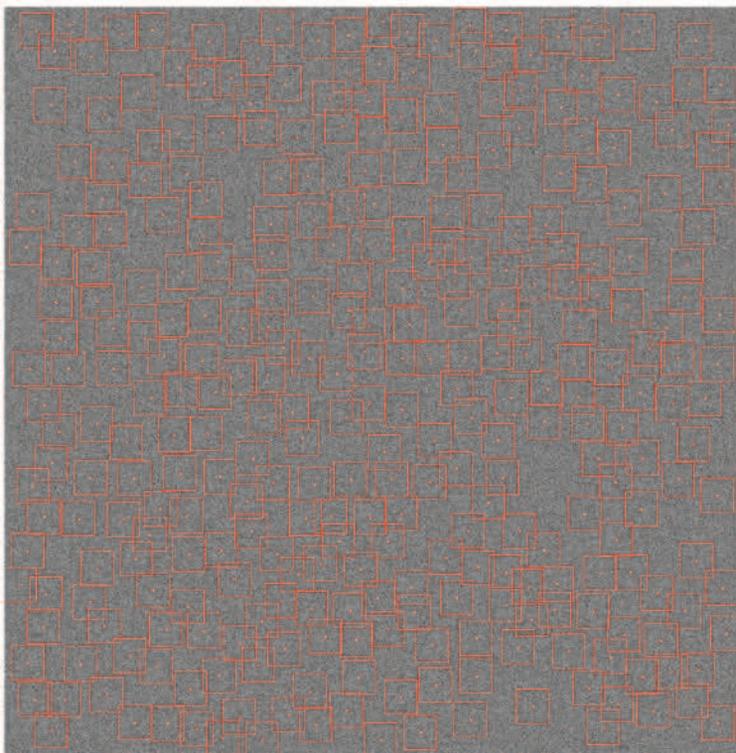

**D** Verification

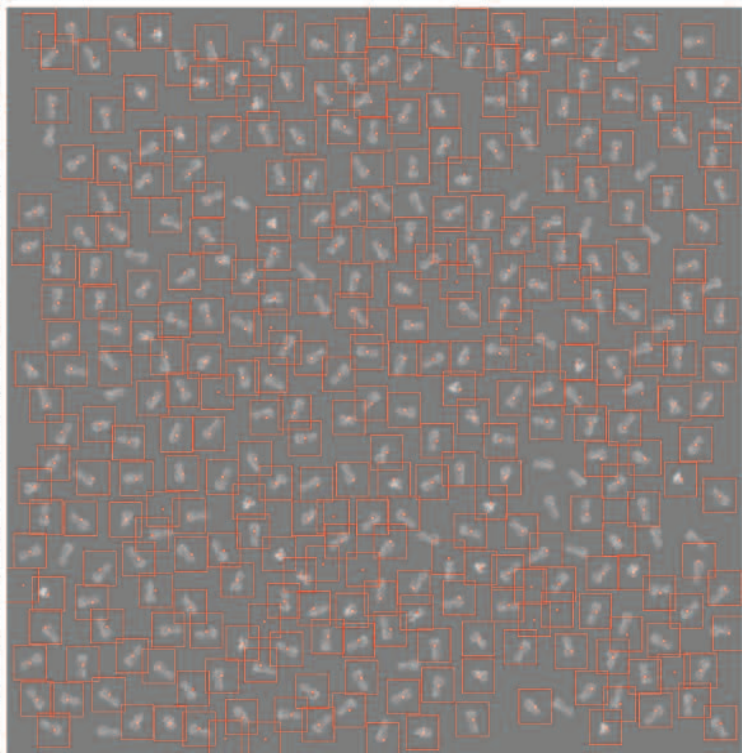

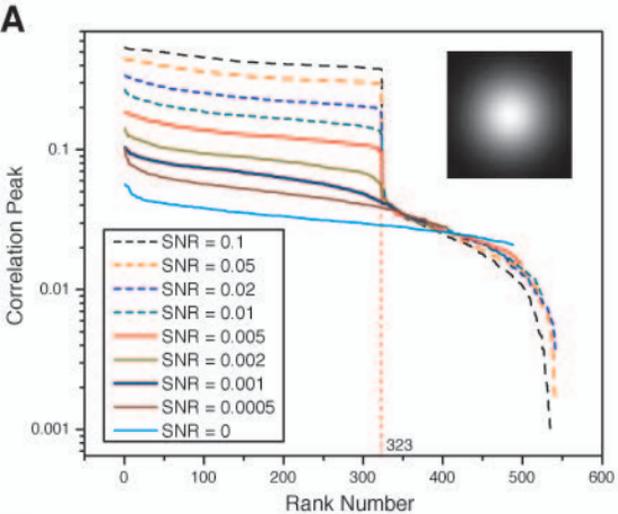

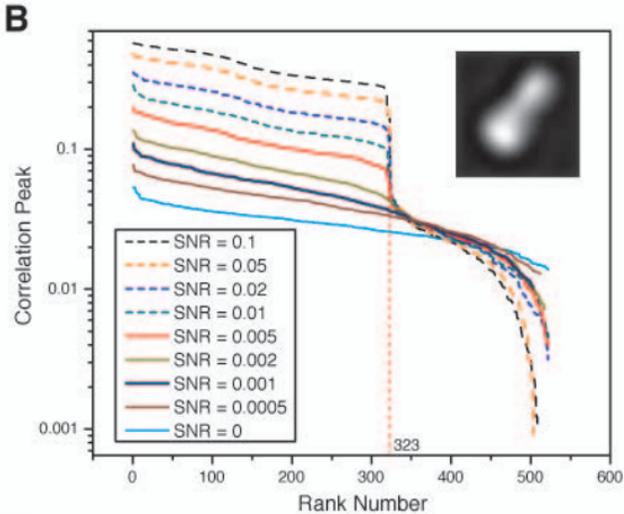

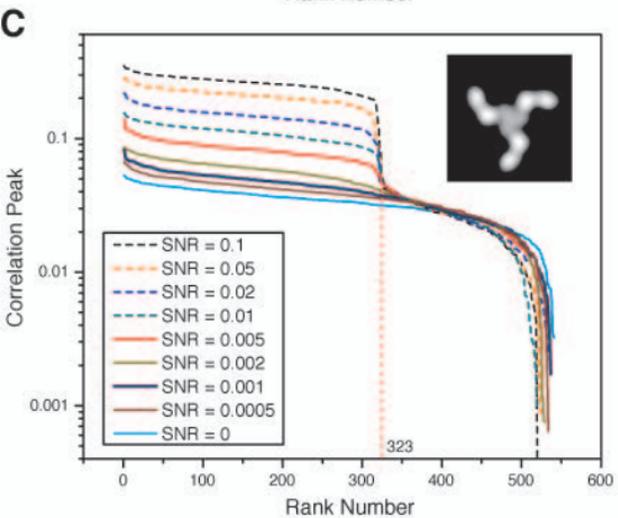

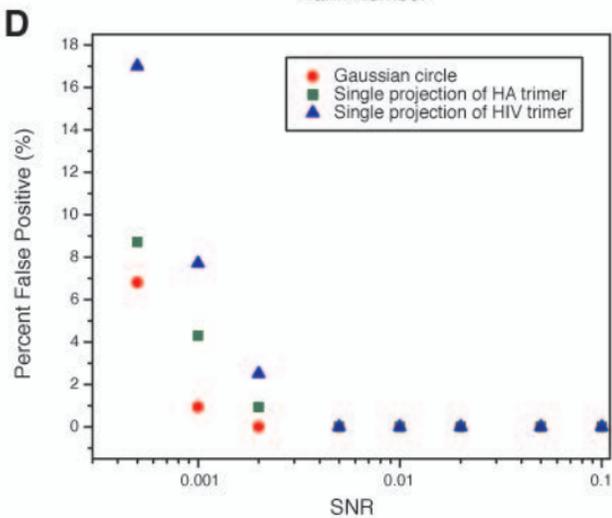

Blind Test                                    Verification

**A**
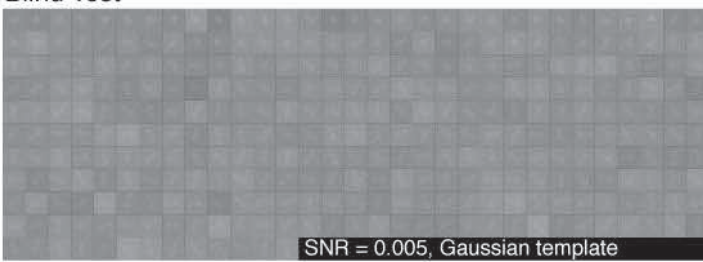 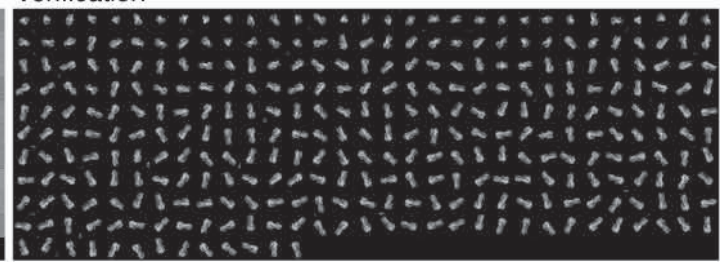
SNR = 0.005, Gaussian template

**B**
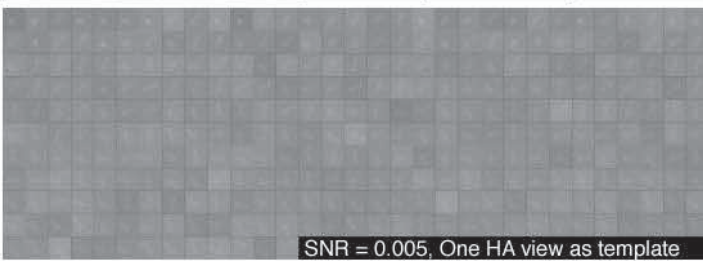 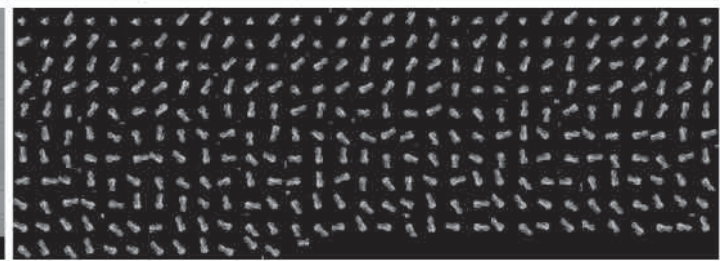
SNR = 0.005, One HA view as template

**C**
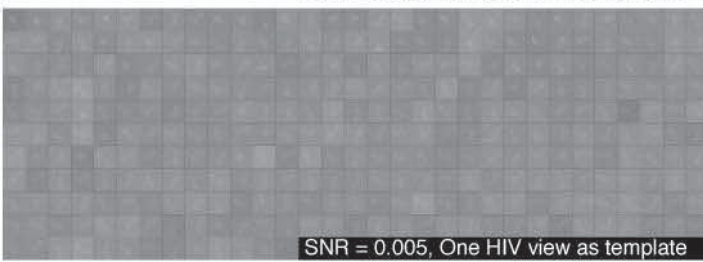 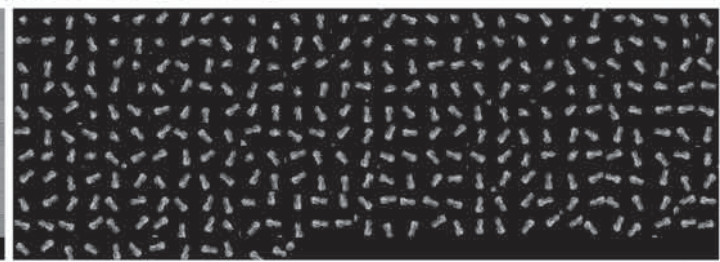
SNR = 0.005, One HIV view as template

**D**
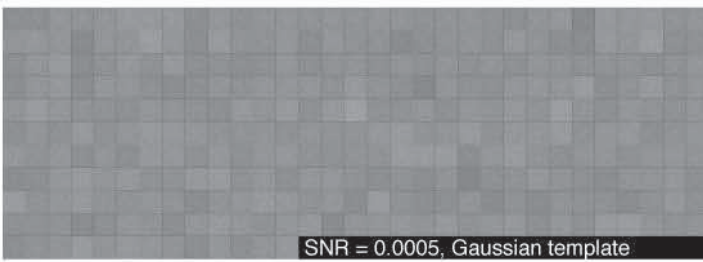 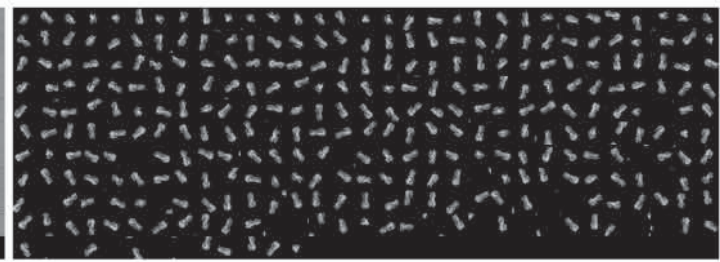
SNR = 0.0005, Gaussian template

**E**
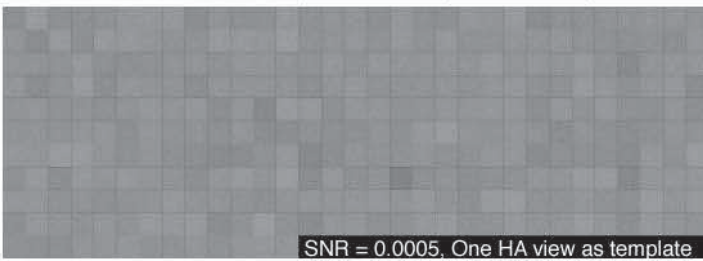 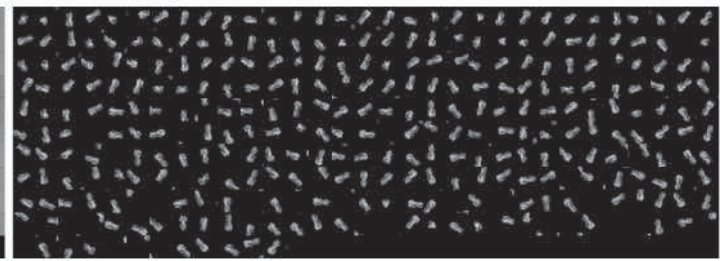
SNR = 0.0005, One HA view as template

**F**
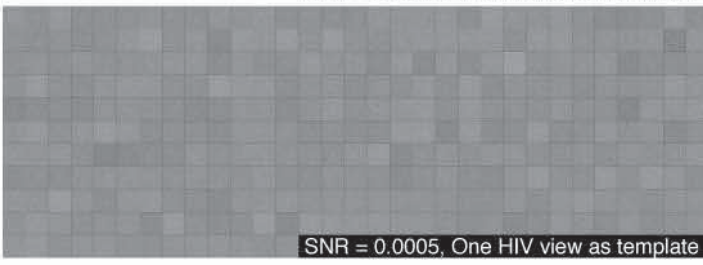 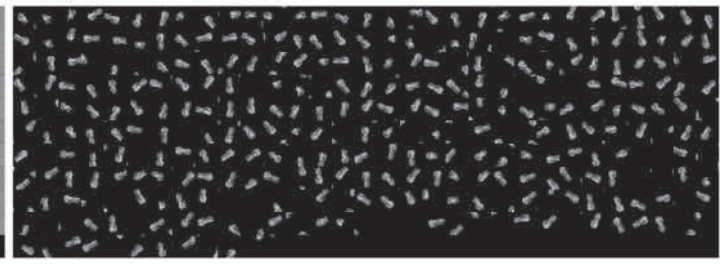
SNR = 0.0005, One HIV view as template

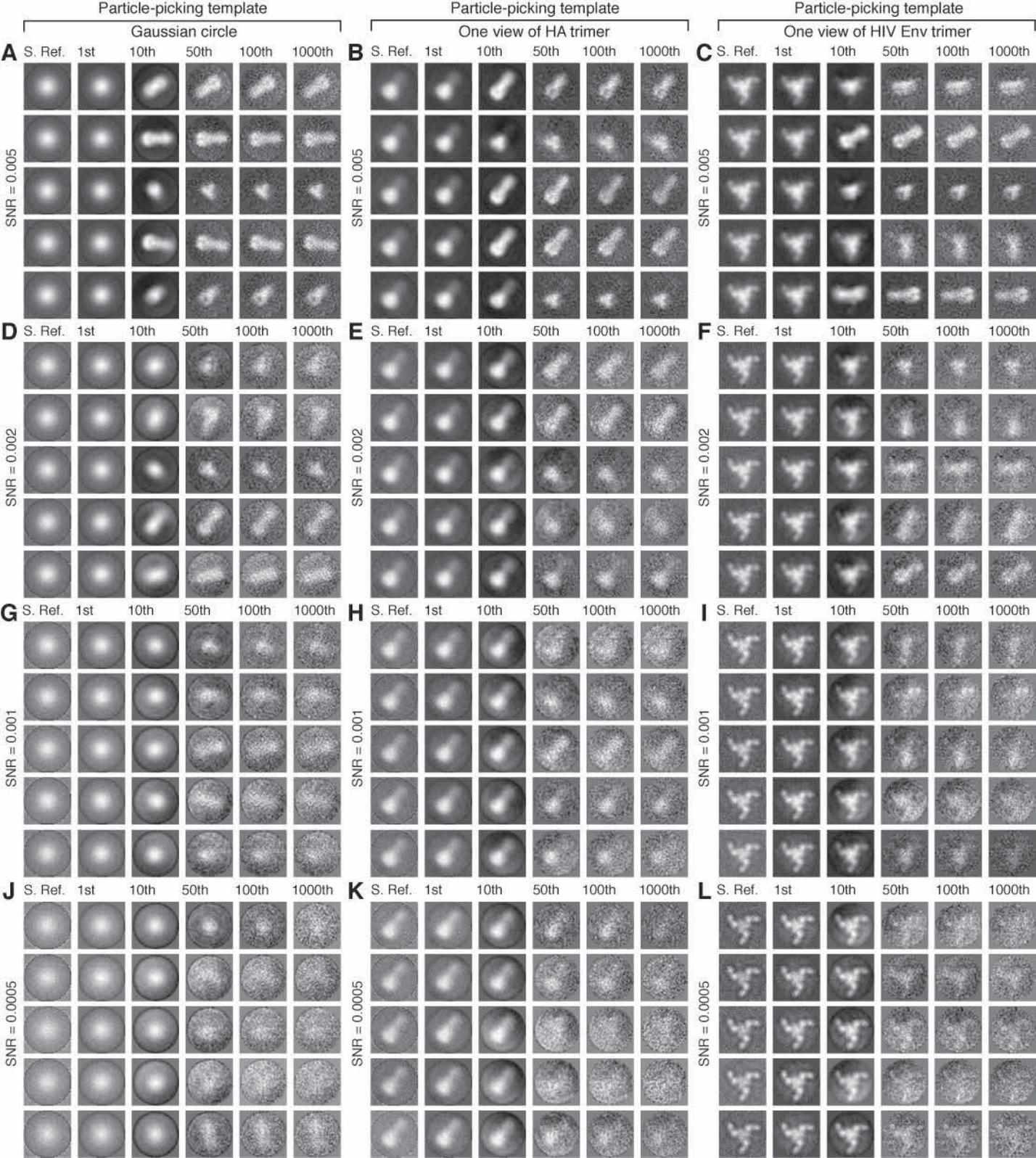